\documentclass[twocolumn, superscriptaddress, unsortedaddress, showpacs,  prl]{revtex4}
\usepackage{amsmath}
\usepackage{graphicx}

\usepackage{natbib}
\usepackage[withbib, all]{authorindex}

\begin{document}
\title{Imaging spin properties using spatially varying magnetic fields}
\author{V. P. Bhallamudi}
\affiliation{Department of Electrical and Computer Engineering, The Ohio State University, Columbus, Ohio 43210, USA}
\author{ A. J. Berger}
\affiliation{Department of Physics, The Ohio State University, Columbus, Ohio 43210, USA}
\author{D. E. Labanowski}
\affiliation{Department of Electrical and Computer Engineering, The Ohio State University, Columbus, Ohio 43210, USA}
\author{D. Stroud}
\affiliation{Department of Physics, The Ohio State University, Columbus, Ohio 43210, USA}
\author{P. C. Hammel\email{hammel@mps.ohio-state.edu}}
\affiliation{Department of Physics, The Ohio State University, Columbus, Ohio 43210, USA}

\date{\today}

\begin{abstract}
We present a novel method to image spin properties of spintronic systems using the spatially confined field of a scanned micromagnetic probe, in conjunction with existing electrical or optical  {\em global} spin detection schemes. It is thus applicable to all materials systems susceptible to either of those approaches. The proposed technique relies on numerical solutions to the spin diffusion equation in the presence of spatially varying fields to obtain the local spin response to the micromagnetic probe field. These solutions also provide insight into the effects of inhomogeneities on Hanle measurements.
\end{abstract}


\maketitle

Spin electronics is a rich and rapidly growing field that integrates magnetic and nonmagnetic materials to enable a variety of spin polarized transport phenomena \cite{Zutic:2004p477, Awschalom:2007}.  Much progress has been made in understanding injection of spins into non-magnetic semiconductors, the manipulation of those spins within the semiconductor, and both local and global detection of the spin states \cite{Jedema:2002p1359,   Crooker:2005p1461, Lou:2007p121,  Furis:2007p479}. Several recent results have brought us closer to electrically-controlled room-temperature spintronic devices \cite{Sasaki:2010p1497,  Sasaki:2010p1408,  Pi:2010p1502, Huang:2008p1463, appelbaum_electronic_2007, Koo:2009p1412, Dash:2009p1462, wunderlich_spin_2010, jansen:InjModDopedSilicon.PRB2010}.  However, much is still not well understood, particularly regarding the mechanisms that degrade spin polarization,  in the often complex magnetic environment that prevails in spintronic devices. Spin lifetimes inferred from measurements are often much shorter than theoretical expectations and the influence of impurities, inhomogeneities and other departures from ideal structures remain poorly understood.

\begin{figure}[ht]
\center{\includegraphics[width=0.8\linewidth]{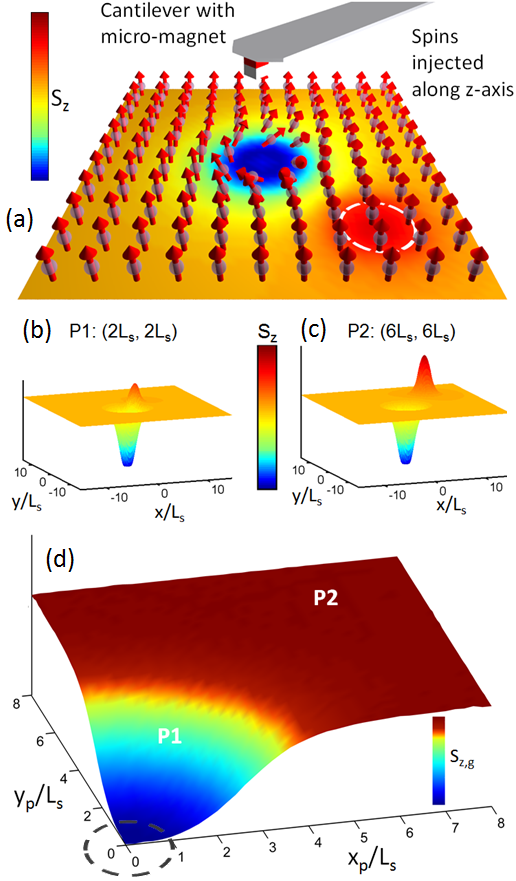}}
\caption{[Color online] Calculated vector magnetization demonstrating imaging of local properties using global measurements: (a) Measurement geometry of our proposed experiment. Embedded in a uniform plane is a small region (dotted line) in which $\tau_s$ is five times larger than throughout the remaining area. The 2D sample is uniformly pumped with spins. A probe is positioned above the sample at a distance $z_p$ and its $xy$-coordinates are scanned to obtain the image presented in panel (d).  (b,c) Spatial map of $S_z (x, y)$ when the enhanced $\tau_s$ region is found at $(2L_s, 2L_s)$ and $(6L_s,6L_s)$ relative to the tip.  (d) $S_{z,g}$, the globally integrated $S_z$ over the entire sample as a function of probe co-ordinates ($x_{p}, y_{p}$).}
	\label{fig:tau_s imaging}
\end{figure}

A technique which could spatially resolve the local variations of spin density and other key parameters, such as spin lifetime, in multicomponent spintronic devices would have enormous impact.  Here we show that local information about the spin properties can be encoded in the variation of the global spin polarization in response to a spatially scanned magnetic field from a micromagnetic probe.   In analogy with electrostatic tip-induced scattering in a two-dimensional electron gas \cite{Westervelt:ScanTip2DEG.Nature2001}  the local magnetization is modified to an extent determined by both the vector field of the micromagnetic probe and the local spin properties. This local perturbation is detected by established global spin polarization measurement techniques (such as spin photoluminescence or non-local voltage) which, in themselves, need not provide spatially resolved information, hence the method is applicable to any materials system (whether optically active or inactive).  A schematic of our proposed imaging technique is shown in Fig. \ref{fig:tau_s imaging}(a). As a magnetic dipole is scanned above the sample the dependence of a globally detected spin signal on  position of the probe provides a map of the sample's spin properties.

Spin dynamics in nonmagnetic materials are given by
\begin{equation}
	\frac{\partial{\bf S}}{\partial t}
     = D_s\nabla^2{\bf S}
     + \gamma{\bf B}\times {\bf S}
     - \frac{{\bf S}}{\tau_s}
     + {\bf G},
	\label{eq:diffeq}
\end{equation}
where $\bf S$ is the spin density, $D_s$ is the spin diffusion constant, $\gamma = g\mu_B/\hbar$ is the gyromagnetic ratio, ${\bf B}$  is the total magnetic field experienced by the spins, $\tau_s$ is the spin relaxation time, and $\bf G$ represents the spin generation term, e.g. irradiation with circularly polarized light. $\bf S$ is a function of time $t$ and spatial position ${\bf r} = (x, y)$ for the 2D sample that we consider.

The spatially dependent response of the spin polarization to the magnetic field of the probe tip can be calculated by means of a numerical technique we have developed for solving the spin diffusion equation in the presence of spatially varying vector magnetic fields.  Vector precession is an added complexity which is not present in the case of (scalar) charge deflection by a scanned gate.  Our technique gives steady state spatial maps of spin density as a function of the location of the micromagnetic probe over the sample. The numerical solution is obtained from the time-dependent form of the diffusion equation using Euler's iterative method (please see supplementary material for further details).  Though we do not consider electric fields nor take ${\bf B}$, ${\bf G}$ and $\tau_s$ to be time-dependent these situations are straightforwardly handled by our approach.

We consider spins injected uniformly into the sample by some means (uniformity is not a necessary condition). The injected spins and the scanned dipole are assumed to be oriented along $\bf{\hat z}$. This sample has a small localized region in which the spin lifetime is five times longer than the rest of the sample.  Panels (b) and (c) of Fig. \ref{fig:tau_s imaging} show spatial maps of $S_z$, the component of the spin density  parallel to the injected spin direction, for two positions of the probe relative to the lifetime inhomogeneity.  Panel (d) presents the image of $S_{z, g}$, the integral of $S_z$ over the entire sample, as a function of the position of dipole over the sample. The region of slower relaxation rate is clearly distinguished from the rest of the sample.

The sensitivity of spin polarization to the position of the probe arises from the interrelationship of the two mechanisms that alter spin polarization: spin relaxation and dephasing of spins in an ensemble as they precess about the spatially inhomogeneous magnetic field vectors. Fields parallel and perpendicular have contrasting impacts since magnetic field parallel to the injected spin direction will tend to maintain this orientation and maximize spin polarization, while transverse components cause precession, thus dephasing and a reduction of spin polarization. This competition between the various components of the magnetic field is well captured by a quantity $\theta_B^2=\gamma^2 B_\perp^2\tau_s^2/(1 + \gamma^2 B_\|^2\tau_s^2)$, which governs the average projection of the spins along the $\parallel$-axis (e.g., for uniform systems, the steady-state value of $S_{\parallel}$ is proportional to $(1 + \theta_B^2)^{-1}$).

\begin{figure}[b]
	 \center{\includegraphics[width=0.9\linewidth]{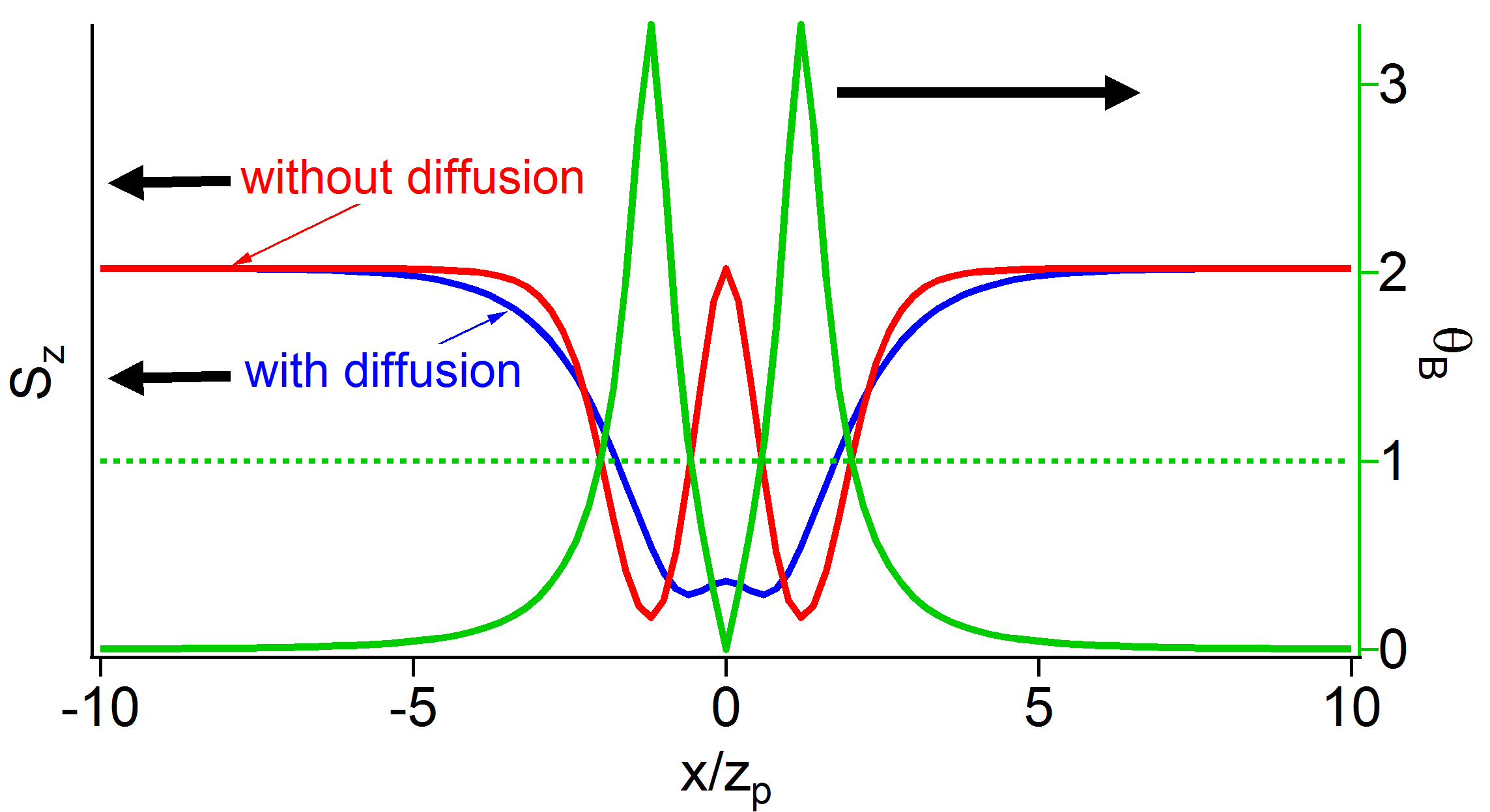}}
	\caption{[Color online] (a) Steady state spin density $S_z(x, y)$ for $y=0$ in the presence of the field of a dipole for  $D_s = 0$ (red) and $D_s\tau_s/z_p^2 = 1$ (blue). The dipole $m_z \hat{\bf z}$ [here $\mu_0 m_z = 10 \pi z_p^3)$]  is located at ${\bf r} =  (0, 0, z_p)$ and uniform injection is assumed.  The green curve shows the spin dephasing factor $\theta_B(x)$; note that $\theta_B = 1 \implies S_z = 0.5$ for $D_s = 0$.}
	\label{fig:thetaBSz}
\end{figure}
We are interested in spatially varying fields arising from a micromagnetic tip. The dipole field arising from a magnetic moment ${\bf m} = m_z \hat{\bf z}$ located at (0, 0, $z_p$) over a sample in the $xy$-plane results in a spatial variation of spin density $S_z$ shown in Fig.~\ref{fig:thetaBSz}  (left axis) along with $\theta_B$ (right axis). Here we assume a sample with uniform $\tau_s$ and uniform injection of spins oriented parallel to $\hat{\bf z}$ and approximate the probe field by ${\bf B}_{\rm dip}(x, y) = \frac{\mu_0}{4\pi}[(3{\bf  R}({\bf m}\cdot{\bf R}) -{\bf m}R^2]/R^{5}$, where ${\bf R} = x \hat{\bf x} + y \hat{\bf y} + z_p \hat{\bf z}$  and $\mu_0$  is the permeability of free space.  Fig.~\ref{fig:thetaBSz} shows $S_z$ both with and without diffusion. The anticorrelation between  $S_z$ and $\theta_B$ is particularly evident when diffusion is neglected because a given spin experiences a single magnetic field throughout its lifetime. Diffusion smears out the sharp features, and if the minima in $S_\parallel$ (for $D_s = 0$) occur closer than a spin-diffusion length $L_s$ away from the central peak, then the peak will collapse leaving a localized volume of suppressed spin density.

This local suppression of the spin polarization by the probe tip field provides the basis for scanned probe imaging of a spintronic sample: the impact of this tip field is determined by local sample properties. The image contrast in Fig.~\ref{fig:tau_s imaging}(d) results from the fact that long-lived spins are more strongly dephased because they experience the tip field for a longer time before being replaced by fresh, well-oriented spins.  Thus, the globally averaged spin density of the sample will be more strongly suppressed when the dipole is over the region of longer $\tau_s$. The size of this effect depends on the ``global'' size over which the spin polarization is integrated.

\begin{figure}[ht]
\center{\includegraphics[width=0.9\linewidth]{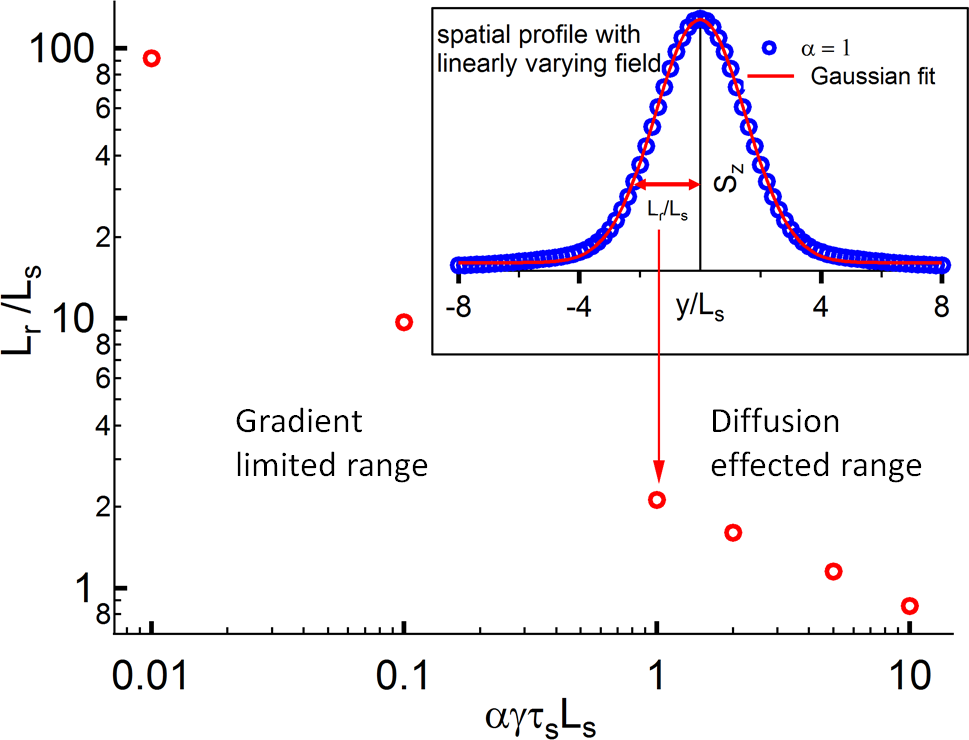}}
	\caption{[Color online] The resolution length scale $L_r$ as a function of magnetic field gradient $\alpha$, where length is scaled by the diffusion length $L_s$, and field is in the units of $(\gamma\tau_s)^{-1}$. (inset) Variation of $S_z$ along the $y$-axis in the presence of a field of the form  ${\bf B} = \alpha (y\hat{\bf y} - z\hat{\bf z})$. A Gaussian fit to the numerically simulated data $S_z$ is shown. The width of the Gaussian may be considered to be the resolution for a given gradient.}
	\label{fig:linear field gradient}
\end{figure}

To estimate the spatial resolution of our imaging technique, we consider the effect of a linear  magnetic field gradient: ${\bf B} = \alpha (y \hat{\bf y} - z \hat{\bf z})$ (see Fig. \ref{fig:linear field gradient}). The injected spins are uniformly oriented along $\hat{\bf z}$.  The resulting variation of $S_\parallel$ along the $y$-axis, for $\alpha = (\gamma \tau_s L_s)^{-1}$, where $(\gamma\tau_s)^{-1}$ is the half-width of the conventional Lorentzian Hanle curve \cite{meier_optical_1984}, is presented in the inset of Fig.~\ref{fig:linear field gradient}.  A Hanle curve is simply a plot of the dependence of spin polarization on a transverse magnetic field.  In the absence of diffusion, spins at different positions along the $y$-axis will sample different fields, resulting in a mapping of the conventional Hanle curve onto the spatial domain. Thus, spins will exist primarily in the region where $|B| \leq (\gamma\tau_s)^{-1}$, i.e.,  where $|y|\leq (\gamma \tau_s \alpha ) ^{-1}$. The location of the surviving spin magnetization can be spatially shifted by adding a uniform offset to the gradient field.

A measure of the resolution may be obtained from the width of a Gaussian fit, $L_r$, to this curve. Fig.~\ref{fig:linear field gradient} shows this width as a function of the applied gradient.  The spatial resolution $L_r$ decreases in inverse proportion to the gradient until $L_r \lesssim L_s$ below which its rate of decreases slows. This situation is reminiscent of Magnetic Resonance Imaging (MRI) where spatial information is encoded into field or frequency shifts through magnetic field gradients \cite{Callaghan:microMRI}. As in MRI, diffusion degrades resolution, though in the current approach it can be finer than the spin diffusion length if one uses strong enough gradients, as seen in numerical simulations shown in Fig. \ref{fig:linear field gradient}.  Furthermore, it should be noted that diffusion lengths for many spintronic systems are of the order of a micron, especially at room temperatures. Field gradients exceeding 10 G/nm are achievable with rare-earth micromagnetic tips \cite{r:tobaccomosaicvirus.pnas.2009} such as are used in Magnetic Force Resonance Microscopy \cite{s:apl91, r:esr, h:MagHandbook} indicating that high spatial resolution is possible with this technique. In the case of imaging with a dipole tip, the gradients (and the regions of suppressed magnetization shown in Fig.~\ref{fig:thetaBSz}) cannot be smaller than the probe-sample separation, $z_p$, so for large $z_p$ this will set the resolution.

Understanding the impact of microscopic inhomogeneities on spin transport is a centrally important issue that calls for spatially resolved studies.  Particularly important are inhomogeneities due to the stray fields arising from imperfect ferromagnetic injectors used for electrical injection \cite{Dash:2009p1462,  AwoAffouda:2009p798, jansen:roughFMinterface.arXiv, jansen:InjModDopedSilicon.PRB2010}.  Spin lifetime is typically obtained from the width of Hanle curves, but such global measures of spin polarization are sensitive to inhomogeneities, so microscopic approaches are needed to discern the various mechanisms that can affect the shape of a Hanle curve.

\begin{figure}[ht]
	  \center{\includegraphics[width=0.9\linewidth]{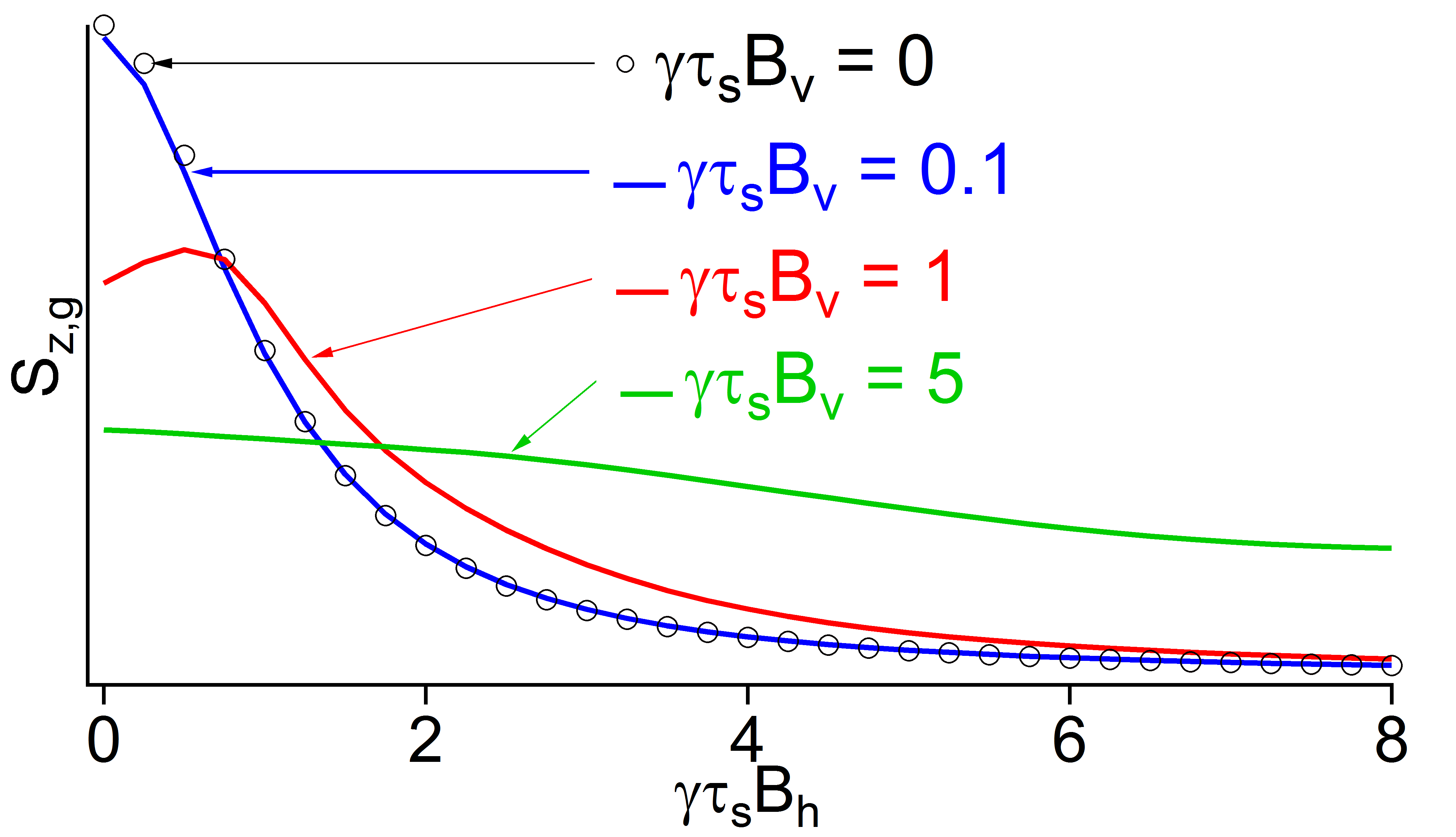}}
	\caption{[Color online] Global Hanle curves for a sample with randomly spatially varying magnetic field, $ {\bf B}_r  = {\cal N}_x(0, B_v)\hat{\bf x} +  {\cal N}_y(0, B_v)\hat{\bf y} +  {\cal N}_z(0, B_v)\hat{\bf z}$. Such fields may be expected in electrical injection devices with rough injectors.  The injected spins are assumed to be along the $x$-axis in this simulation.}
	\label{fig:random field}
\end{figure}

Non-uniformity of the interface and complex domain structures can cause the injected spin population to experience significantly different magnetic fields, and lose spin-information on time scales much shorter than spin lifetime due to dephasing. Our numerical method is capable of simulating the impact of magnetic fields with random spatial variation.  Fig.~\ref{fig:random field} shows calculated Hanle curves in the presence of random magnetic fields. We consider a field ${\bf B}_r  = {\cal N}_x(0, B_v) \hat{\bf x} +   {\cal N}_y(0, B_v) \hat{\bf y} +   {\cal N}_z(0, B_v) \hat{\bf z}$ where we sample normal distributions of mean zero and variance $B_v$ for each field component at every spatial point. Fig.~\ref{fig:random field} shows results for three values of $B_v: \, \ll (\gamma\tau_s)^{-1},  \, = (\gamma\tau_s)^{-1}, \,\mbox{and} \, >(\gamma\tau_s)^{-1}$. We see that the maxima can be shifted and, as in the last case, the curves can be significantly broadened.

\begin{figure}[ht] \center{\includegraphics[width=0.9\linewidth]{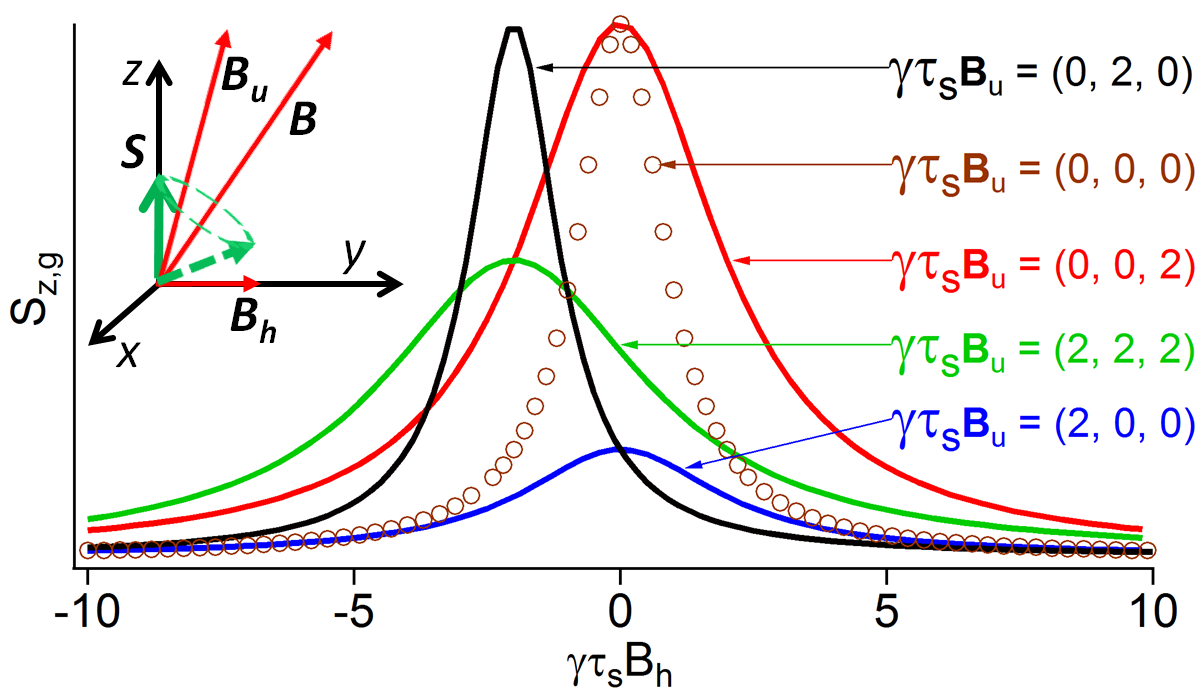}}
	\caption{[Color online] Global Hanle curves for various cases of spatially uniform magnetic fields, $ {\bf B}_u$, applied in addition to the swept Hanle field, $ \bf{B}_h$. These curves are valid for all values of $D_s$ since globally averaged spin density is insensitive to diffusion (see Supplementary Information), when the sample dimensions are much greater than $L_s$. }
	\label{fig:uniform field}
\end{figure}

It is instructive to consider, in turn, the effects of each magnetic field component on the Hanle curve, as is done in Fig. \ref{fig:uniform field}.  For an applied Hanle field $\textbf{B}_h = B_h \bf{\hat{y}}$, the half-width at half-maximum $B_{1/2}$ of a Hanle curve in the presence of an additional uniform field $\textbf{B}_u = B_x \hat{\textbf{x}} + B_y \hat{\textbf{y}} + B_z \hat{\textbf{z}}$ is given by $\gamma  \tau_s B_{1/2} = \sqrt{1+\gamma^2\tau_s^2(B_x + B_z)^2}$\,. A transverse field $B_x$ reduces $S_{z,g}(B_h = 0)$ (blue curve) and $B_h \sim B_x$ is required to significantly reduce $S_{z,g}$ further. A non-zero $B_z$ (red curve) on the other hand will not  reduce $S_{z,g}(B_h = 0)$ but will broaden the Hanle curve, since  $B_h \sim B_z$ is  required to significantly increase the spin precession cone angle. A field $B_y$ (parallel to $B_h$) shifts the peak of the Hanle curve, since $S_{z,g}$ is maximized when the {\it total} transverse field is zero (black curve). The results of Fig. \ref{fig:random field} can be considered a superposition of the shifting and broadening seen in Fig. \ref{fig:uniform field}. Clearly these effects can confound the determination of spin lifetimes from Hanle widths. The detailed features of the stray field generated by a rough ferromagnetic injector will depend on several characteristics including roughness, saturation magnetization, domain size, and thickness. Regardless of these details, stray fields of the order kiloGauss are readily achievable a few nanometers away from a rough injector and these can substantially affect the spin polarization.  In fact this has been experimentally observed in silicon devices \cite{jansen:roughFMinterface.arXiv}.

In summary, we have proposed a new technique for imaging spin properties of spatially inhomogeneous samples that achieves applicability to a wide variety of materials by relying on proven spin polarization detection techniques such as electrical and optical detection.  The localized information is obtained by detecting the modification to the local spin density by the confined magnetic field of a magnetic dipole scanned over the sample.  To this end, we have developed a method for simulating spin density in a medium with spatial or temporal inhomogeneities of the magnetic field, lifetime, or diffusion constant, regardless of injection or detection technique.  As in MRI, this technique has a spatial resolution inversely proportional to the gradient of the magnetic field down to length scales comparable to the spin diffusion length. Below this, resolution improves more slowly with increasing gradient. Our numerical analysis of spin transport emphasizes the importance of  experimental characterization of microscopic inhomogeneity, and of models that incorporate these phenomena. In particular, inhomogeneities can confound the interpretation of spin polarization and lifetime measurements, as in Hanle curves.  Microscopic imaging of spin properties of real-world devices will be essential for improving their spin fidelity and functionality.

We gratefully acknowledge enlightening discussions with Ron Jansen. This work was supported by National Science Foundation through the Materials Research Science and Engineering Center at The Ohio State University (DMR-0820414) and Department of Energy, Office of Science (DE-FG02-03ER46054).

\appendix
\section*{Supplementary Information}

Spin dynamics in nonmagnetic materials are described by
\begin{equation}
	\frac{\partial{\bf S}}{\partial t}
     = D_s\nabla^2{\bf S}
     + \gamma{\bf B}\times {\bf S}
     - \frac{{\bf S}}{\tau_s}
     + {\bf G},
	\label{eq:diffeq}
\end{equation}
where $\bf S$ is the spin density, $D_s$ is the spin diffusion constant, $\gamma = g\mu_B/\hbar$ is the gyromagnetic ratio, ${\bf B}$  is the total magnetic field experienced by the spins, $\tau_s$ is the spin relaxation time, and $\bf G$ represents the spin generation term e.g., irradiation with circularly polarized light. $\bf S$ is a function of time $t$ and spatial position ${\bf r} = (x, y)$ for the 2D sample that we consider.  While all of the terms can be functions of $t$ and $\bf r$, we will consider ${\bf B, ~ G}$ and $\tau_s$ which are independent of $t$, and $D_s$ will be independent of both space and time. The formalism can be  readily extended to include electric fields,  spin-orbit effects, hyperfine coupling and a third spatial dimension for a more comprehensive treatment.  Also, we are usually interested in the {\it steady-state} solution $\partial{\bf S}/\partial t = 0$ and  $S_\parallel$, the parallel component of $\bf S$. We will refer to ``parallel'' and ``perpendicular'' with respect to the injected spin direction. Experimentally $S_\parallel$ is the most commonly measured quantity. However, it should be noted that we can evaluate any component of the spin in our simulations.

If ${\bf B}$ and $\tau_s$ are {\it position-independent}, the steady-state differential equation can be solved analytically by Fourier transform.  In the steady state, the Fourier-transformed equation for the spin density ${\bf S}({\bf k})$ takes the form
\begin{align}
	 -k^2 D_s&{\bf S}({\bf k})
     + \gamma{\bf B}\times{\bf S}({\bf k})
     - {\bf S}({\bf k})/\tau_s + {\bf G}({\bf k}) = 0 \nonumber \\
	\implies &{\bf S}({\bf k})
      = [(k^2 D_s + \tau_s^{-1})I-{\cal B}]^{-1}{\bf G}({\bf k})
	\label{eq:FT soln}
\end{align}
where ${\bf S}({\bf k})$ and ${\bf G}({\bf k})$ are vectors. $I$ is the $3 \times 3$ unit matrix. We introduce the 3 $\times$ 3 matrix
\[{\cal B} = \gamma\left(\begin{array}{ccc}
0 	 & B_z 	& -B_y\\
-B_z & 0		& B_x \\
B_y  & -B_x 	& 0  \end{array} \right), \]
  The real-space spin density is then given by the inverse Fourier transform of $\bf{S}(\bf{k})$.

An important case experimentally is given by the globally averaged spin polarization
\begin{equation}
S_{\parallel,g} = \frac{\int_A S_\parallel({\bf r})\,d^2r}{\tau_s\int_A G_\parallel({\bf r})\,d^2r},
\end{equation}
for ${\bf G} = G_\parallel\hat{\parallel}$ and the area of integration, A, extends over the entire space. Note that $S_{z,g} \propto S_{\parallel} (k=0)$ and thus from Eq.~\ref{eq:FT soln} is independent of $D_s$ and is given by
\begin{equation}
	S_{\parallel,g} =  \frac{1}{1 + \theta_B^2}
	\label{eq:matrixsol}
\end{equation}
where $\theta_B$ is an effective {\it dephasing factor} given by
\begin{equation}
	\theta_B^2
     =\frac{\gamma^2 B_\perp^2\tau_s^2}
           {1 + \gamma^2 B_\|^2\tau_s^2}
	\label{eq:thetaB}
\end{equation}
Spins precess in a cone whose opening half-angle is that between the total vector magnetic field and the injected spin orientation. In combination with this precession, the continuous injection of spins causes a distribution of phases, relative to $\hat{\parallel}$, weighted by the spin lifetime. A parallel field reduces the cone opening angle resulting in a spin ensemble more aligned with the injected spin direction, while $B_\perp$ has the opposite effect. The dephasing factor $\theta_B$ (representing the competition between parallel and transverse fields) describes the projection of spins along the injection axis and the behavior of $S_{\parallel,g}$.

The experimentally relevant situation will usually involve spatially varying fields. If ${\cal B}$ is position-dependent and $D_s \neq 0$, Eq. \ref{eq:diffeq} cannot be solved analytically in real or Fourier space for steady-state. We have found that solving the original time-dependent equation numerically using an Euler method works well for this problem.  In this approach, we start from some initial condition, ${\bf S}({\bf r}, t = 0)$,  then iterate in time using a time step $\Delta t$, according to the relation
\multlinetaggap=0.0pt
\begin{eqnarray}
	{\bf S}({\bf r}, t + \Delta t) &=& {\bf S}({\bf r}, t) + \Delta t \times
	 \left[ \rule[-3mm]{0mm}{4mm}D_s\nabla^2{\bf S}({\bf r}, t)\right.   \nonumber \\
     &+& \left.\gamma{\bf B}({\bf r})\times {\bf S}({\bf r}, t) 
     - \frac {{\bf S}({\bf r}, t)}{\tau_s}
     + {\bf G}({\bf r}) \right] \nonumber
\end{eqnarray}
The Laplacian is evaluated numerically on a spatial grid of points separated by a suitable distance $\Delta x$, and is approximated as
\begin{equation}
\nabla^2{\bf S}({\bf r}, t) \sim  \frac{1}{(\Delta x)^2}\sum_{\hat{\delta}}[{\bf S}({\bf r} +  \Delta x \hat{\delta}, t) - {\bf S}({\bf r}, t)]
\end{equation}
 where $\hat{\delta}$ represents the unit vectors along the spatial grid. We iterate until $\bf S$ does not change appreciably with time. We have verified that the analytical results from the analytical expressions derived above are reproduced by this method.  Several experimental global and spatially localized Hanle curves measured under varying conditions\cite{ Furis:2007p479, Jedema:2002p1359} can also be qualitatively reproduced by this method.


\begin{thebibliography}{10}

\bibitem{Zutic:2004p477}
I. {\v Z}uti{\'c}, J. Fabian, and S.~D. Sarma, Rev. Mod. Phys.  (2004).

\bibitem{Awschalom:2007}
D.~D. Awschalom and M.~E. Flatte, Nature Phys. {\bf 3},  153  (2007).

\bibitem{Jedema:2002p1359}
F. Jedema {\it et~al.}, Nature {\bf 416},  713  (2002).

\bibitem{Crooker:2005p1461}
S. Crooker {\it et~al.}, Science {\bf 309},  2191  (2005).

\bibitem{Lou:2007p121}
X. Lou {\it et~al.}, Nature Phys. {\bf 3},  197  (2007).

\bibitem{Furis:2007p479}
M. Furis {\it et~al.}, New J. Phys. {\bf 9},  347  (2007).

\bibitem{Sasaki:2010p1497}
T. Sasaki {\it et~al.}, IEEE Trans. Magn. {\bf 46},  1436  (2010).

\bibitem{Sasaki:2010p1408}
T. Sasaki {\it et~al.}, Appl. Phys. Lett. {\bf 96},  122101  (2010).

\bibitem{Pi:2010p1502}
K. Pi {\it et~al.}, Phys. Rev. Lett. {\bf 104},  187201  (2010).

\bibitem{Huang:2008p1463}
B. Huang and I. Appelbaum, Phys. Rev. B {\bf 77},  165331  (2008).

\bibitem{appelbaum_electronic_2007}
I. Appelbaum, B. Huang, and D.~J. Monsma, Nature {\bf 447},  295  (2007).

\bibitem{Koo:2009p1412}
H.~C. Koo {\it et~al.}, Science {\bf 325},  1515  (2009).

\bibitem{Dash:2009p1462}
S.~P. Dash {\it et~al.}, Nature {\bf 462},  491  (2009), supplementary
  information.

\bibitem{wunderlich_spin_2010}
J. Wunderlich {\it et~al.}, 1008.2844  (2010).

\bibitem{jansen:InjModDopedSilicon.PRB2010}
R. Jansen {\it et~al.}, Phys. Rev. B {\bf 82},  241305  (2010).

\bibitem{Westervelt:ScanTip2DEG.Nature2001}
M. Topinka {\it et~al.}, Nature {\bf 410},  183  (2001).

\bibitem{meier_optical_1984}
F. Meier and B.~P. Zakharchenya, {\em Optical Orientation} ({North-Holland},
  Amsterdam, 1984).

\bibitem{Callaghan:microMRI}
P.~T. Callaghan, {\em Principles of Nuclear Magnetic Resonance Microscopy}
  (Clarendon Press, Oxford, 1991).

\bibitem{r:tobaccomosaicvirus.pnas.2009}
C.~L. Degen {\it et~al.}, Proc.\ Natl.\ Acad.\ Sci.\ USA {\bf 106},  1313
  (2009).

\bibitem{s:apl91}
J.~A. Sidles, Appl.\ Phys.\ Lett.\ {\bf 58},  2854  (1991).

\bibitem{r:esr}
D. Rugar, C.~S. Yannoni, and J.~A. Sidles, Nature {\bf 360},  563  (1992).

\bibitem{h:MagHandbook}
P.~C. Hammel and D.~V. Pelekhov,  in {\em Handbook of Magnetism and Advanced
  Magnetic Materials}, edited by H. Kronm\"uller and S. Parkin (John Wiley \&
  Sons, Ltd., New York, NY, 2007), Vol.~5, Chap.~4.

\bibitem{AwoAffouda:2009p798}
C. Awo-Affouda {\it et~al.}, Appl. Phys. Lett. {\bf 94},  102511  (2009).

\bibitem{jansen:roughFMinterface.arXiv}
S.~P. Dash {\it et~al.}, arXiv:1101.1691v1 [cond-mat.mes-hall] (2011).

\end{thebibliography}
\end{document}